\documentclass{PoS}

\newcommand{\im}{\mbox{Im }}
\newcommand{\re}{\mbox{Re }}

\newcommand{\mpi}{M_\pi}
\newcommand{\mps}{\mpi^2}

\newcommand{\PV}{\mathcal{P.P.}}

\title{Once subtracted Roy-like dispersion relations and a precise
analysis of $\pi\pi$ scattering data}

\ShortTitle{Precise analysis of $\pi\pi$ scattering data}

\author{\speaker{R. Garc\'\i{}a-Mart\'\i{}n}
\thanks{We thank the organizers for creating the nice scientific
atmosphere of the event and the Spanish research contracts
PR27/05-13955-BSCH, FPA~2004-02602, UCM-CAM~910309
and BFM~2003-00856 for partial financial support. RGM's work
is partially supported by the European commission MRTN FLAVIAnet
[MRTN-CT-2006035482].}\\
        Dpto. F\'\i{}sica Te\'orica II, Universidad Complutense, Madrid, Spain {\em and}\\
        Groupe de Physique Th\'eorique, Institut de Physique Nucl\'eaire (IN2P3-CNRS), Orsay, France.\\
        E-mail: \email{garcia@ipno.in2p3.fr}}
\author{R. Kami\'nski\\
        Institute of Nuclear Physics, Polish Academy of Sciences, Cracow, Poland.\\
        E-mail: \email{Robert.Kaminski@ifj.edu.pl}}
\author{J. R. Pel\'aez\\
        Dpto. F\'\i{}sica Te\'orica II, Universidad Complutense, Madrid, Spain\\
        E-mail: \email{jrpelaez@fis.ucm.es}}
\author{F. J. Yndur\'ain\\
        Departamento de F\'\i{}sica Te\'orica C-XI, Universidad Aut\'onoma de Madrid, Cantoblanco, Spain.}

\abstract{We report our progress on the data analysis of $\pi\pi$ scattering
data in terms of
Forward Dispersion Relations (FDR), as well as
Roy equations (RE) and their once-subtracted
counterpart, GKPY equations. The first part of the analysis
consists of independent
fits to the different $\pi\pi$ channels.
The GKPY equations provide a more stringent consistency check
for the parametrizations of the $S0$-wave data in the region from 400 to 1100 MeV, 
In the second part we present our preliminary analysis where the fits are constrained
to satisfy all dispersion
relations within errors, including the new GKPY Eqs., thus providing a very precise and
model independent description of
data using just analyticity, causality and crossing.}

\FullConference{International Workshop on Effective Field Theories: from the pion to the upsilon \\
		February 2-6 2009\\
		Valencia, Spain}

\begin{document}

\section{Introduction}

A precise knowledge of pion-pion scattering is of interest 
since it provides a test of Chiral Perturbation Theory (ChPT)
as well as useful information about 
quark masses and the chiral condensate \cite{Gasser:1983yg}. It is
also of interest to establish the properties of the $f_0(600)$ or sigma meson
(see talk by J.~R.~Pel\'aez
\cite{JRPelaez} in this conference and references therein).
This reaction, at least in the elastic regime, is also remarkably 
 symmetric in terms of isospin and crossing symmetries.
Unfortunately, the existing experimental information on $\pi\pi$
scattering has many conflicting data sets at intermediate
energies and used to have no data at all close to the interesting threshold region.
For many years this fact has
made it very hard to obtain conclusive results on $\pi\pi$ scattering
at low energies or in the sigma region. However, recent \cite{Batley:2007zz}
and precise experiments
on kaon decays, related to $\pi\pi$ scattering at very low energies, have renewed
the interest on this process.

The dispersive integral formalism is 
model independent, just based on analyticity and crossing,
and relates the $\pi\pi$ amplitude 
at a given energy with an integral over the whole energy range, increasing the precision.
It also provides information
on the amplitude either at energies where data are poor
or in the complex plane.  In addition, 
it makes the parametrization of the data irrelevant once it is included in the integral
and relates different 
scattering channels among themselves. For all these reasons it is well suited to
study the threshold region or the poles in the complex plane associated to resonances.
(For our progress and references on the latter, see
\cite{JRPelaez} in this conference).

Our recent works make use of two complementary 
dispersive approaches, in brief:
\paragraph{Forward Dispersion Relations (FDRs)} 
They are calculated at $t=0$ so that the 
unknown large-t behavior of the amplitude is not needed.  
We consider two symmetric and one asymmetric isospin combinations, 
to cover the isospin basis.  The symmetric ones,
$\pi^0\pi^+$ and $\pi^0\pi^0$, have two subtractions and
can be written as
\begin{equation}
\re F_{00,0+}(s,0)-F_{00,0+}(4\mps,0)=
\frac{s(s-4\mps)}{\pi}\,\PV\int_{4\mps}^\infty ds'
\frac{(2s'-4\mps)\,\im F_{00,0+}(s',0)}{s'(s'-s)(s'-4\mps)(s'+s-4\mps)}
\label{FDR1}
\end{equation}
All contributions 
to their integrands are positive, which makes them very precise.
The antisymmetric isospin combination $I_t=1$ does not require two subtractions:
\begin{equation}
F^{(I_t=1)}(s,0)=
\frac{2s-4\mps}{\pi}\,\PV\int_{4\mps}^\infty ds'
\frac{\im F^{(I_t=1)}(s',0)}{(s'-s)(s'+s-4\mps)}. 
\label{FDR2}
\end{equation}
We have implemented all of them up to $\sqrt{s}\simeq1420$~MeV
\paragraph{Roy Equations (RE) \cite{Roy:1971tc}}
An infinite set of coupled equations
fully equivalent to nonforward 
dispersion relations, plus some $t-s$ crossing symmetry, 
 written in terms of partial waves of definite isospin I and angular momentum $l$.
The complicated left cut contribution is rewritten 
as series of integrals over partial waves in the physical region:
\begin{equation}
\re f^{(I)}_\ell(s)=C_l^{(I)}a_0^{(0)}+{C'_\ell}^{(I)}a_0^{(2)}
+\sum_{I',\ell'}
\PV\int_{4\mps}^\infty ds' 
K_{\ell\ell'}^{II'}(s';s)\,\im f^{(I')}_{\ell'}(s').
\label{Roy}
\end{equation}
where the  $C_\ell^{(I)}$, ${C'_\ell}^{(I)}$ constants (actually,
first order polynomials in $s$) and
$K_{\ell\ell'}^{II'}$ kernels
are known.  
In practice, the calculation is truncated at $\ell<2$ and at some cutoff
energy $s_0$. The
$\ell\geq2$ waves and the high energy are treated as input.
RE are well suited to study poles of resonances but are limited
to $\sqrt{s}\leq 8\mpi \simeq 1120$~MeV.
At present, we have implemented them up to $\sqrt{s}\simeq2m_K$.
%In previous works we have implemented them up to $\sqrt{s}\simeq 2M_K$,
%a more complete analysis up to $\sqrt{s}=1115$~MeV is sketched here
%and will be fully published in an upcoming paper.

In the last decade the use of RE has regained interest with several aims: to improve the precision
of scattering data, to discard spurious solutions, to test ChPT, or to use ChPT
to obtain the subtraction constants at low energies, which can be recast in
terms of the scattering lengths $a^{(0)}_{0}$ and $a^{(2)}_{0}$,
and to obtain precise
predictions on $\pi\pi$ scattering. 
In addition it is of interest for the lightest scalar resonance, but that will
be detailed in J.R. Pel\'aez's talk in this conference \cite{JRPelaez}.
In particular 
a series of RE analysis in \cite{Bern} using $\pi\pi$ data parametrizations
for the $\ell>2$ waves and above 800 MeV for the rest,
as well as some Regge input, was performed 
with and without ChPT constraints. 
The latter provided, $a^{(0)}_{0}=0.220\pm0.005\,M_\pi^{-1}$
and $a^{(2)}_{0}=-0.0444\pm0.0010\,M_\pi^{-1}$, an extremely precise claim,
together with predictions for other scattering
lengths and the $S$- and $P$-wave phase shifts up to 800 MeV.
Some of the input, particularly the Regge theory and the
$D$ waves, was questionable~\cite{critica}, but it certainly seems to have
a very small influence in the threshold region of the scalar
waves~\cite{Caprini:2003ta}. Also in these period, 
the Krakow-Paris \cite{Kaminski:2002pe}
and Paris \cite{DescotesGenon:2001tn} groups have performed
other RE analysis. The former resolved the long-standing ambiguity,
discarding the so-called "up"' solution, including in their analysis
a study using polarized target data. The latter checked the
calculation in \cite{Bern} and claimed an small discrepancy
in the Olsson sum rule. Recently, our group has also made a series of
works on a dispersive analysis of $\pi\pi$ scattering 
that we review in the next section.

In section 3 we advance some results
on our derivation of an additional set of Roy-like dispersive
equations, which are obtained by starting with a once-subtracted
dispersion relation. They impose additional constraints at intermediate
energies, and allow us to reduce the uncertainty of the analysis.

\section{Overview of the analysis}
The approach we have followed throughout a series of
works~\cite{Kaminski:2006qe} can be summarized as follows:
\begin{enumerate}
\item We first obtain simple fits to data for each $\pi\pi$ scattering
partial wave (the so called {\em Unconstrained Fits to Data}, or {\em UFD08}
for short). These fits are uncorrelated, therefore they can be very easily
changed when new, more precise data become available. This has
actually happened, for example,
in~\cite{Yndurain:2007qm}, where we have included the newest, very precise
$K_{l_4}$ data~\cite{Batley:2007zz}.
At different stages of our approach we have also fitted
Regge theory to $\pi\pi$ high energy data, and as our precision
was improving, we have improved some 
of the UFD fits with more flexible parametrizations.
\item Then, these UFD08 are checked against FDR, several sum rules,
and Roy's Equations.
Surprisingly, some widely used data parametrizations fail to
satisfy the FDR or some of the sum rules. Thus we keep those
parametrizations that are in better agreement with the FDR.
\item Finally, we impose these dispersive relations in the previous
fits as additional constraints. These new {\em Constrained Fits
to Data} ({\em CFD08} for short) are much more precise and reliable
than the UFD08 set, being consistent with analyticity, unitarity, crossing,
etc. The price to pay is that now all the waves are correlated.
\end{enumerate}

In order to quantify how well the dispersion relations are satisfied,
we define six quantities $\Delta_i$ as the difference between the left and right
sides of each dispersion relation
in Eqs.~(\ref{FDR1}), (\ref{FDR2}) and (\ref{Roy}),
whose uncertainties we call $\delta\Delta_i$.
Next, we define the average discrepancies
\begin{equation}
\bar{d}_i^2\equiv\frac{1}{\hbox{number of points}}
\sum_n\left(\frac{\Delta_i(s_n)}{\delta\Delta_i(s_n)}\right)^2,
\end{equation}
where the values of $s_n$ are taken at intervals of 25 MeV. 
Note the similarity with an averaged $\chi^2/(d.o.f)$
and thus $\bar{d}_i^2\leq1$
implies fulfillment of the corresponding dispersion relation within errors.
In Table 1 we show the average discrepancies of the UFD08 
for each FDR, up to two different energies, and each RE up to $\sim2m_K$. 
Although the total average discrepancy of the UFD08 set is
practically one, they can be clearly improved in the high energy region
of the antisymmetric FDR and in the scalar isospin-2 RE.
This is actually done 
in the CDF08 set, which is obtained by minimizing:
\begin{equation}
\chi^2\equiv
\left\{\bar{d}^2_{00}+\bar{d}^2_{0+}+\bar{d}^2_{I_t=1}+\bar{d}^2_{S0}+\bar{d}^2_{S2}+\bar{d}^2_P\right\}
W
+\bar{d}^2_I+\bar{d}^2_J+\sum_i\left(\frac{p_i-p_i^{\rm exp}}{\delta p_i}\right)^2.
\end{equation}
where $p_i^{exp}$ are all the parameters of the different UFD08
parametrization for each wave or Regge trajectory, 
thus ensuring the data description, and
$d_I$ and $d_J$ are the discrepancies for a couple of crossing
sum rules. The weight $W=9$ was estimated from
the typical number of degrees of freedom needed to describe
the shape of the dispersion relations.

\begin{table}
\begin{tabular}{l|cc||cc}
&\multicolumn{2}{c||}{ Unconstrained Fits to Data (UFD08)}&\multicolumn{2}{c}{Constrained Fits to Data (CFD08)}\\
\hline
&$s^{1/2}\leq 932\,$MeV&$s^{1/2}\leq 1420\,$MeV&$s^{1/2}\leq 992\,$MeV&$s^{1/2}\leq 1420\,$MeV\\
$\pi^0\pi^0$ FDR& 0.12 & 0.29 & 0.13 & 0.31\\
$\pi^+\pi^0$ FDR& 0.84 & 0.86 & 0.83 & 0.85\\
$I_{t=1}$ FDR& 0.66 & 1.87 & 0.13 & 0.70\\
\hline
&\multicolumn{2}{c||}{$s^{1/2}\leq 992\,$MeV}&\multicolumn{2}{c}{$s^{1/2}\leq 992\,$MeV}\\
Roy eq. S0&\multicolumn{2}{c||}{0.54}&\multicolumn{2}{c}{0.23}\\
Roy eq. S2&\multicolumn{2}{c||}{1.63}&\multicolumn{2}{c}{0.25}\\
Roy eq. P &\multicolumn{2}{c||}{0.74}&\multicolumn{2}{c}{0.002}\\
\hline 
\end{tabular}
\caption{Average discrepancies $\bar d^2$ of the UFD08 and CFD08 for each 
FDR and RE. On average, the UFD are 
consistent with dispersion relations. Note the remarkable CFD08 consistency.}
%\vspace*{.1cm}}
\end{table}

From the Table it is clear that the CFD set satisfies remarkably 
well all dispersion relations within uncertainties, and 
hence can be used directly and inside the Olsson sum rule
to obtain the following precise determination
{\em from data}: $a^{(0)}_{0}=0.223\pm0.009\,m_\pi^{-1}$ and
$a^{(2)}_{0}=-0.0444\pm0.0045\,m_\pi^{-1}$.
This is in remarkable agreement with the predictions of RE
and ChPT of \cite{Bern}.
Nevertheless, the agreement is fairly good only up to roughly 450 MeV,
but from that energy up to 800 MeV those predictions 
deviate from our data analysis. We should stress that
we are nevertheless talking about a disagreement of a few degrees at most
and would affect the determination of the sigma mass by tens
of MeV at most, which is a remarkable improvement compared with
the situation just a few years ago and the
huge and extremely conservative uncertainties quoted in the PDG
for the $\sigma$ mass and width, of hundreds of MeV.

The other waves are of less relevance for this conference and we
comment them very briefly, since the details can be found
in~\cite{Kaminski:2006qe}. The best determination of threshold
parameters is obtained by using the CFD08 set inside appropriate
sum rules~\cite{Kaminski:2006qe}. 
In brief, we agree with \cite{Bern} in the $P$-wave scattering length,
but find disagreements of 2 to 3 standard deviations in the
P-wave slope, and also in some $D$-wave parameters. 

In summary the CFD08 set provides a model independent and very precise
description
of the $\pi\pi$ scattering data consistent with analyticity and crossing. 

\section{Work in progress: GKPY equations and 400 to 1100 MeV region}
\begin{figure}
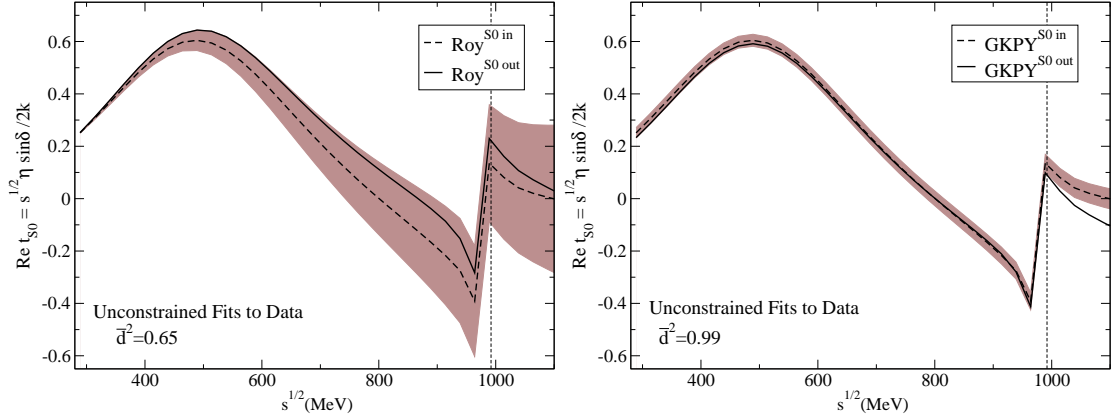

  \centering
  \includegraphics[height=5.5cm]{udf-roys0.eps}
  \includegraphics[height=5.5cm]{udf-gkpys0.eps}
  \caption{Above 450~MeV the uncertainty in the standard
$S0$-wave Roy eqs. (left) is much larger than in the once-subtracted
GKPY equations (right). We show results for the new UFD S0 wave parametrization.}
  \label{fig:errors}
\end{figure}

\begin{figure}
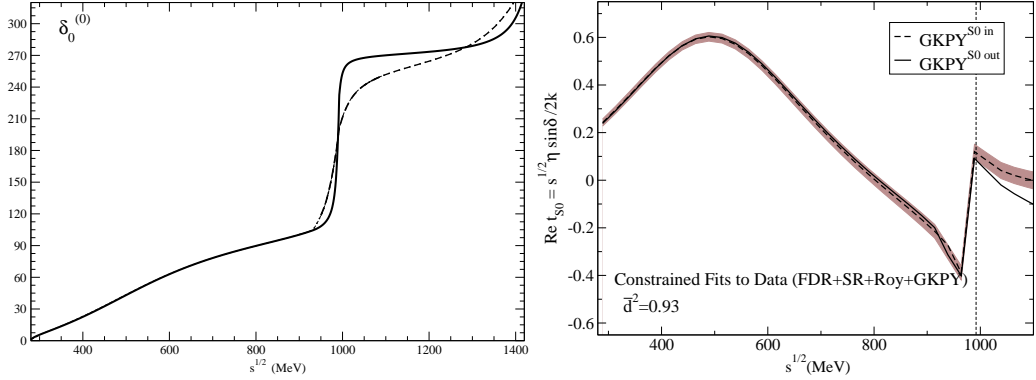

  \centering
\vspace{.2cm}
  \includegraphics[height=5.0cm]{cdf-s0.eps}
  \includegraphics[height=5.0cm]{cdf-gkpys0.eps}
  \caption{Left: The improved matching only changes the $S0$-wave above
932~MeV. Right: agreement with GKPY eqs. is improved above 850~MeV
in the preliminary CFD with improved matching and GKPY.}
  \label{fig:s0}
\end{figure}

The precision of the analysis sketched in the previous section
in the $\sqrt{s}>400$~MeV region is limited by the big uncertainties
that Roy's equations propagate at high energies, in particular those
coming from the {\it experimental} uncertainty in $a^{(2)}_0$. By starting with a
once-subtracted dispersion relation we arrive at the following set of
Roy-like equations (GKPY for brevity):
\begin{equation}
  \label{eq:gkpy}
  \re f^{(I)}_\ell(s)=D_\ell^{(I)}+\sum_{I',\ell'}
  \PV\int_{4\mps}^\infty ds' 
  \tilde K_{\ell\ell'}^{II'}(s';s)\,\im f^{(I')}_{\ell'}(s').
\end{equation}
in which the $\tilde K_{\ell\ell'}^{II'}$ kernels and $D_\ell^{(I)}$
subtraction constant are known and will be explicitely given in
an upcoming work. It is important to note that, whereas the subtraction
constants in RE constitute a first-order polynomial in $s$, they
are here just combinations of $a_0^{(0)}$ and $a_0^{(2)}$. Therefore,
the propagation of uncertainties (see Fig.~\ref{fig:errors}) is as follows:
for $\sqrt{s}<400$~MeV RE have smaller uncertainties, but for
$\sqrt{s}>450$~MeV GKPY equations have a significantly smaller error
band. This enables further improvements in our fits for the $S0$-wave
in the 400 to 1100 MeV region. In fact, the level of precision attained
in this region forces us to refine our fits by means of
an improved matching between the $S0$-wave
parametrizations at low and intermediate energies, which
occurs at 932~MeV, imposing also continuity in the first derivative.
Preliminary results for the phase shifts can be seen in Fig.~\ref{fig:s0} (left),
to compare with
that of the previous fit (dashed line). For clarity, we do not provide
the data points, which are nevertheless reasonably well described by
both parametrizations when taking into account the experimental errors.

Finally, we have obtained a new CFD using this improved UFD as 
the starting point, and adding the GKPY equations as an additional
constraint. The results obtained (see plot on the right in
Fig.~\ref{fig:s0}) are very similar to the
previous CFD, but GKPY fulfillment is improved. 

\section*{Acknowledgments}

This work is dedicated to the memory of Prof. Yndur\'ain who recently passed 
away but participated decisively in the main discussions and initial calculations
of this work.


\begin{thebibliography}{99}
%\cite{Gasser:1983yg}
\bibitem{Gasser:1983yg}
  J.~Gasser and H.~Leutwyler,
  %``Chiral Perturbation Theory To One Loop,''
  Annals Phys.\  {\bf 158}, 142 (1984).
  %%CITATION = APNYA,158,142;%%
  
  %\cite{Batley:2007zz}
\bibitem{Batley:2007zz}
  J.~R.~Batley {\it et al.}  [NA48/2 Collaboration],
  %``New high statistics measurement of K(e4) decay form factors and pi pi
  %scattering phase shifts,''
  Eur.\ Phys.\ J.\  C {\bf 54} (2008) 411.
  %%CITATION = EPHJA,C54,411;%%
    %\cite{Pislak:2001bf}
%\bibitem{Pislak:2001bf}
  S.~Pislak {\it et al.}  [BNL-E865 Collaboration],
  %``A new measurement of K+(e4) decay and the s-wave pi pi scattering  length
  %a(0)(0),''
  Phys.\ Rev.\ Lett.\  {\bf 87} (2001) 221801
  [arXiv:hep-ex/0106071].
  %%CITATION = PRLTA,87,221801;%%

%\cite{JRPelaez}
\bibitem{JRPelaez}
  J.~R.~Pel\'aez, R.~Garcia-Martin and R.~Kaminski,
  on this conference.
  %%CITATION = ARXIV:xxxx.xxxx;%%

  %\cite{Roy:1971tc}
\bibitem{Roy:1971tc}
  S.~M.~Roy,
  %``Exact integral equation for pion pion scattering involving only physical
  %region partial waves,''
  Phys.\ Lett.\  B {\bf 36}, 353 (1971).
  %%CITATION = PHLTA,B36,353;%%

%\cite{Bern}
\bibitem{Bern}
%\bibitem{Colangelo:2001df}
  G.~Colangelo, J.~Gasser and H.~Leutwyler,
  %``pi pi scattering,''
  Nucl.\ Phys.\  B {\bf 603} (2001) 125
%  [arXiv:hep-ph/0103088].
  %%CITATION = NUPHA,B603,125;%%
  %\cite{Ananthanarayan:2000ht}
%\bibitem{Ananthanarayan:2000ht}
  B.~Ananthanarayan, G.~Colangelo, J.~Gasser and H.~Leutwyler,
  %``Roy equation analysis of pi pi scattering,''
  Phys.\ Rept.\  {\bf 353} (2001) 207
 % [arXiv:hep-ph/0005297].
  %%CITATION = PRPLC,353,207;%%
  
\bibitem{critica}
%\cite{Pelaez:2005wj}
%\bibitem{Pelaez:2005wj}
  J.~R.~Pelaez and F.~J.~Yndurain,
  %``Chiral-dispersive calculations of pion pion scattering confront
  %experiment,''
  Nucl.\ Phys.\ Proc.\ Suppl.\  {\bf 164} (2007) 93;
%  [arXiv:hep-ph/0510216].
  %%CITATION = NUPHZ,164,93;%%
  %\cite{Pelaez:2003eh}
%\bibitem{Pelaez:2003eh}
%  J.~R.~Pelaez and F.~J.~Yndurain,
  %``On the precision of chiral-dispersive calculations of pi pi scattering,''
  Phys.\ Rev.\  D {\bf 68} (2003) 074005
%  [arXiv:hep-ph/0304067].
  %%CITATION = PHRVA,D68,074005;%%

  %\cite{Caprini:2003ta}
\bibitem{Caprini:2003ta}
  I.~Caprini, G.~Colangelo, J.~Gasser and H.~Leutwyler,
  %``On the precision of the theoretical predictions for pi pi scattering,''
  Phys.\ Rev.\  D {\bf 68} (2003) 074006
%  [arXiv:hep-ph/0306122].
  %%CITATION = PHRVA,D68,074006;%%

  %\cite{Kaminski:2002pe}
\bibitem{Kaminski:2002pe}
  R.~Kaminski, L.~Lesniak and B.~Loiseau,
  %``Elimination of ambiguities in pi pi phase shifts using crossing symmetry,''
  Phys.\ Lett.\  B {\bf 551}, 241 (2003)
%  [arXiv:hep-ph/0210334].
  %%CITATION = PHLTA,B551,241;%%
  
  %\cite{DescotesGenon:2001tn}
\bibitem{DescotesGenon:2001tn}
  S.~Descotes-Genon, N.~H.~Fuchs, L.~Girlanda and J.~Stern,
  %``Analysis and interpretation of new low-energy pi pi scattering data,''
  Eur.\ Phys.\ J.\  C {\bf 24}, 469 (2002)
 % [arXiv:hep-ph/0112088].
  %%CITATION = EPHJA,C24,469;%%

%\cite{Kaminski:2006qe}
\bibitem{Kaminski:2006qe}
  R.~Kaminski, J.~R.~Pelaez and F.~J.~Yndurain,
  %``The pion-pion scattering amplitude. III: Improving the analysis with
  %forward dispersion relations and Roy equations,''
  Phys.\ Rev.\  D {\bf 77} (2008) 054015;
 % [arXiv:0710.1150 [hep-ph]].
  %%CITATION = PHRVA,D77,054015;%%
  %\cite{Kaminski:2006yv}
%\bibitem{Kaminski:2006yv}
 % R.~Kaminski, J.~R.~Pelaez and F.~J.~Yndurain,
  %``The pion pion scattering amplitude. II: Improved analysis above anti-K K
  %threshold,''
  Phys.\ Rev.\  D {\bf 74} (2006) 014001
 % [Erratum-ibid.\  D {\bf 74} (2006) 079903]
 % [arXiv:hep-ph/0603170].
  %%CITATION = PHRVA,D74,014001;%%

  %\cite{Yndurain:2007qm}
\bibitem{Yndurain:2007qm}
R.~Garcia-Martin and J.~R.~Pelaez,  F.~J.~Yndurain, 
  %``Experimental status of the pi pi isoscalar S wave at low energy: f0(600)
  %pole and scattering length,''
  Phys.\ Rev.\  D {\bf 76} (2007) 074034
%  [arXiv:hep-ph/0701025].
  %%CITATION = PHRVA,D76,074034;%%
  %\cite{Pelaez:2004vs}
%\bibitem{Pelaez:2004vs}
  J.~R.~Pelaez and F.~J.~Yndurain,
  %``The pion pion scattering amplitude,''
  Phys.\ Rev.\  D {\bf 71} (2005) 074016;
%  [arXiv:hep-ph/0411334].
  %%CITATION = PHRVA,D71,074016;%%
  %\cite{Pelaez:2003ky}
%\bibitem{Pelaez:2003ky}
%  J.~R.~Pelaez and F.~J.~Yndurain,
  %``Regge analysis of pion pion (and pion kaon) scattering for energy  s**(1/2)
  %> 1.4-GeV,''
  Phys.\ Rev.\  D {\bf 69} (2004) 114001
%  [arXiv:hep-ph/0312187].
  %%CITATION = PHRVA,D69,114001;%%



\end{thebibliography}
\end{document}